\documentclass{SPW9_proc}

\begin{document}

\title{Stokes vectors and Minkowski spacetime: Structural parallels}

\author[1,2,*]{J.O. Stenflo}

\affil[1]{Institute for Particle Physics and Astrophysics, ETH Zurich, CH-8093 Zurich}
\affil[2]{Istituto Ricerche Solari Locarno (IRSOL), Via Patocchi, CH-6605 Locarno, Switzerland}
\affil[*]{\textit{Email:} stenflo@astro.phys.ethz.ch}

\runningtitle{Stokes vectors and Minkowski spacetime}
\runningauthor{J.O. Stenflo}

\firstpage{1}

\maketitle

\begin{abstract}
Polarized radiative transfer is most effectively described in terms of Stokes 4-vectors
and Mueller $4\times 4$ matrices. Here we show that the Stokes
formalism has astounding structural parallels with the formalism used for
relativity theory in Minkowski spacetime. The structure
and symmetry properties of the Mueller matrices are the same as those for the
matrix representations of the electromagnetic tensor and the Lorentz
transformation operator. Comparison between these various matrices
shows that the absorption terms 
$\eta_k$ in the Mueller matrix directly correspond to the electric
field components $E_k$ in the electromagnetic tensor and the Lorentz
boost terms $\gamma_k$ in the Lorentz transformation matrix, while the
anomalous dispersion terms $\rho_k$ correspond to the magnetic field
components $B_k$ and the spatial rotation angles $\phi_k$. In a
Minkowski-type space spanned by the Stokes $I,Q,U,V$ parameters, the
Stokes vector for 100\,\%\ polarized light is a null vector living on the surface of null cones,
like the energy-momentum vector of massless particles in ordinary
Minkowski space. Stokes vectors for partially polarized light live
inside the null cones like the momentum vectors for massive
particles. In this description the depolarization of Stokes vectors
appears as a ``mass'' term, which has its origin in a symmetry breaking caused by the incoherent
superposition of uncorrelated fields or wave packets, without the need
to refer to a ubiquitous Higgs field as is done in particle physics. The rotational
symmetry of Stokes vectors and Mueller matrices is that of spin-2
objects, in contrast to the spin-1 nature of the electromagnetic
field. The reason for this difference is that the Stokes objects have substructure: they are 
formed from bilinear tensor products between spin-1 objects, the Jones
vectors and Jones matrices. The governing physics takes place at the
substructure level. 
\end{abstract}

\section{Introduction}\label{sec:intro}
The different disciplines in physics often evolve in isolation from
each other, developing their own formalisms and terminologies. This
difference in concepts, language, and notations hampers the mutual
understanding and deepens the divisions. It has the
effect of obscuring the circumstance that the different disciplines
may have much more in common than meets the eye, and that much of what
seems to be different is of cultural and not of intrinsic origin. 

In the present paper we highlight the non-trivial circumstance that
the mathematical symmetries and structures that govern the Mueller
matrices in polarization theory are the same as those of the electromagnetic tensor in Maxwell
theory and the Lorentz transformations in relativistic physics, all of
which obey the same Lorentz algebra. Absorption, electric
fields, and Lorentz boosts are governed by the symmetric part of the
respective transformation matrix, while 
dispersion, magnetic fields, and spatial rotations are represented by
the antisymmetric part. These two aspects can be
unified in terms of complex-valued matrices, where
the symmetric and antisymmetric aspects represent the real and
imaginary parts, respectively. 

We can extend this comparison by introducing a Minkowski metric for a
4D space spanned by the four Stokes parameters $I,Q,U,V$. In this
description Stokes vectors that represent 100\,\%\ polarization
are null vectors, while partial depolarization causes the Stokes
vector to lie inside the null cones like the energy-momentum vectors
of massive particles in ordinary spacetime. This comparison points to an analogy
between depolarization (which can be seen as a symmetry breaking) and
the appearance of mass. Another interesting property is that, in
contrast to electromagnetism and Lorentz transformations, Stokes
vectors and Mueller matrices have the rotational symmetry of spin-2
objects because they have substructure: they
are formed from bilinear products of spin-1 objects. Here we try to
expose these potentially profound connections and discuss their
meanings. 

We start in Sect.~\ref{sec:emanalog} by showing the
relations between the Stokes formalism in polarization physics and the
covariant formulation of the Maxwell theory of electromagnetism. After
clarifying the symmetries it is shown how the introduction of
complex-valued matrices leads to an elegant, unified formulation. In
Sect.~\ref{sec:lorentz} we show how the Lorentz transformations with
its boosts and spatial rotations have the same structure. The
introduction of a Minkowski metric for polarization space in
Sect.~\ref{sec:depolmass} reveals a
null cone structure for Stokes vectors that represent fully polarized
light. It further brings out an analogy between depolarization caused by the
incoherent superposition of fields and the appearance of what may be
interpreted as a mass term. In Sect.~\ref{sec:spin2} we highlight the
circumstance that the Stokes vectors and Mueller matrices in
polarization physics have the symmetry of spin-2 objects, because they
have substructure, being formed from bilinear products of vector
objects. Section \ref{sec:conc} summarizes the conclusions.

\section{The electro-magnetic analogy}\label{sec:emanalog}
Let $\vec{S}_\nu$ be the 4D Stokes vector for frequency
$\nu$. Explicitly, in terms of its transposed form (with superscript
$T$) and omitting index $\nu$ for clarity of notation, 
$\vec{S}^T\!\equiv (S_0,\,S_1,\,S_2,\,S_3) \equiv (I,\,Q,\,U,\,V)$. The equation for
the transfer of polarized radiation can then be written as 
\begin{equation}\label{eq:transeq}
{{\rm d}\vec{S}_\nu\over{\rm d}\tau_c}=(\vec{\eta}+\vec{I})\,\vec{S}_\nu
\,-\vec{j}_\nu/\kappa_c\,,
\end{equation}
where $\tau_c$ is the continuum optical depth, $\vec{I}$ is the
$4\times 4$ identity matrix (representing continuum absorption),
$\kappa_c$ is the continuum absorption coefficient, $\vec{j}_\nu$ is
the emission 4-vector, while the Mueller absorption matrix
$\vec{\eta}$ that represents the polarized processes due to the
atomic line transitions is 
\begin{equation}\label{eq:etamat} 
\vec{\eta}=\left(\matrix{\,\,\eta_I &\phantom{-}\eta_Q &\phantom{-}\eta_U&
\phantom{-}\eta_V\,\,\cr \,\,\eta_Q&\phantom{-}\eta_I&
\phantom{-}\rho_V&-\rho_U\,\,\cr \,\,\eta_U&-\rho_V&\phantom{-}\eta_I&\phantom{-}
\rho_Q\,\,\cr \,\eta_V&\phantom{-}\rho_U&-\rho_Q&
\phantom{-}\eta_I\,\, }\right)\,.
\end{equation}
 For a detailed account of Stokes vector polarization theory with its
notations and terminology we refer to the monographs of
\citet{stenflo-book94} and \citet{stenflo-lanlan04}. 

While $\eta_{I,Q,U,V}$ represents the absorption terms for the four
Stokes parameters $I,Q,U,V$, the differential phase shifts, generally
referred to as anomalous dispersion or
magnetooptical effects, are represented by the $\rho_{Q,U,V}$
terms. They are formed, respectively, from the imaginary and the real
part of the complex refractive index that is induced when the atomic
medium interacts with the radiation field. 

We notice that $\vec{\eta}$ can be expressed as the sum of two
matrices: a symmetric matrix that only contains the $\eta$ terms, and
an anti-symmetric matrix that only contains the $\rho$
terms. Antisymmetric matrices represent spatial rotations, as will be
particularly clear in Sect.~\ref{sec:lorentz} when comparing with
Lorentz transformations. 

These symmetries turn out to be identical to the symmetries that
govern both the Maxwell theory for electromagnetism and the Lorentz
transformation of the metric in relativity. Here we will compare these
various fields of physics to cast light on the intriguing
connections. We begin with electromagnetism. 

The analogy between the Stokes formalism and electromagnetism has
previously been pointed out by \citet{stenflo-lanlan81} using a 3+1
(space + time) formulation rather than the covariant formulation in
Minkowski spacetime. It is however only with the covariant formulation that
the correspondences become strikingly transparent. 

The Lorentz electromagnetic force law when written in covariant 4D
form is 
\begin{equation}\label{eq:forcelaw} 
{{\rm d}p^{\,\alpha}\over\!\!{\rm d}\tau}={e\over
  m}\,F^{\,\alpha}_{\phantom{\alpha}\beta}\,\,p^{\,\beta}\,,
\end{equation}
where $p^{\,\alpha}$ are the components of the contravariant energy-momentum
4-vector, $e$ and $m$ are the electric charge and mass of the
particle, and $F^{\,\alpha}_{\phantom{\alpha}\beta}$ are the
components of the electromagnetic tensor, which has the representation
of a $4\times 4$ matrix: 
\begin{equation}\label{eq:fabmat} 
F^{\,\alpha}_{\phantom{\alpha}\beta}\,=\left(\matrix{\,\,0&\phantom{-}E_x&\phantom{-}E_y&
\phantom{-}E_z\,\,\cr \,\,E_x&\phantom{-}0&
\phantom{-}B_z&-B_y\,\,\cr \,\,E_y&-B_z&\phantom{-}0&\phantom{-}
B_x\,\,\cr \,\,E_z&\phantom{-}B_y&-B_x&
\phantom{-}0\,\, }\right)\,.
\end{equation}
 A comparison between Eqs.~(\ref{eq:etamat}) and (\ref{eq:fabmat})
immediately reveals the correspondence between absorption $\eta$ and the electric
$E$ field on the one hand and between anomalous dispersion $\rho$ and the magnetic $B$
field on the other hand: 
\begin{eqnarray}\label{eq:emconnect} 
&\eta_{Q,U,V}\,\,\longleftrightarrow \,\, E_{x,y,z}\nonumber\\
&\,\,\rho_{Q,U,V}\,\,\longleftrightarrow \,\, B_{x,y,z}\,.
\end{eqnarray}

Let us follow up this structural comparison in terms of a unified
description, where absorption and dispersion are combined into a
complex-valued absorption. $\eta$ is proportional to the Voigt
function $H(a,v_q)$, with $\eta_0$ as the proportionality constant, $a$
the dimensionless damping parameter, and $v_q$ the dimensionless
wavelength or frequency parameter. Index $q$, with $q=0,\pm 1$,
indicates the differential shift of the wavelength scale for atomic
transitions with magnetic quantum number $m_{\rm lower}-m_{\rm
  upper}=q$. Similarly $\rho$ is proportional 
to $2F(a,v_q)$, where $F$ is the line dispersion function. 
\begin{eqnarray}\label{eq:etarho} 
&\,\,\,\eta_{I,Q,U,V}=\,\,\eta_0 \,H_{I,Q,U,V}\,,\nonumber\\
&\rho_{Q,U,V}=2\eta_0 \,F_{Q,U,V}\,.
\end{eqnarray}
In the unified description the $H$ and $F$ functions are combined into
the complex-valued 
\begin{equation}\label{eq:hcal}
{\cal H}(a,v_q)\equiv H(a,v_q)-2i\,F(a,v_q)\,, 
\end{equation}
which now represents the building blocks when forming the
corresponding quantities with indices $I,Q,U,V$ to refer to the
respective Stokes parameters. ${\cal H}_{I,Q,U,V}$ can be combined
into the 4-vector 
\begin{equation}\label{eq:hcalvect} 
\vec{{\cal H}}\,\equiv\left(\matrix{\,\,{\cal H}_I\,\,\cr \,\,{\cal H}_Q\,\,\cr \,\,{\cal H}_U\,\,\cr \,\,{\cal H}_V\,\, }\right)\,\equiv\,\left(\matrix{\,\,{\cal H}_0\,\,\cr \,\,{\cal H}_1\,\,\cr \,\,{\cal H}_2\,\,\cr \,\,{\cal H}_3\,\, }\right)\,.
\end{equation}
 From the above follows how ${\cal H}_k$ is related to and unifies the
corresponding absorption and dispersion parameters $\eta_k$ and $\rho_k$: 
\begin{equation}\label{eq:completa}
\eta_k -i\,\rho_k =\,\eta_0\,{\cal H}_k\,. 
\end{equation}

Let us next define the three symmetric matrices $\vec{K}^{(k)}$ and three
antisymmetric matrices $\vec{J}^{(k)}$ through 
\begin{eqnarray}\label{eq:ikj} 
&\vec{K}^{(k)}_{0j}&\!\!\!\equiv\,\, \vec{K}^{(k)}_{j\,0}\,\,\equiv 1\,,\nonumber\\
&\vec{J}^{(k)}_{ij}&\!\!\!\equiv\, -\vec{J}^{(k)}_{j\,i}\equiv-\,\varepsilon_{ij\,k}\,,
\end{eqnarray}
where $\varepsilon_{ij\,k}$ is the Levi-Civita antisymmetric symbol. 
We further define the complex-valued matrix 
\begin{equation}\label{eq:xmatdef}
\vec{T}^{(k)} \equiv \vec{K}^{(k)} -i\,\vec{J}^{(k)} \,. 
\end{equation}
Then the Mueller matrix from Eq.~(\ref{eq:etamat}) becomes 
\begin{equation}\label{eq:etacompmat} 
\vec{\eta} -\eta_I \vec{I} =\,\eta_0\,{\rm Re}\,(\,{\cal
  H}_k\,\vec{T}^{(k)} \,)\,. 
\end{equation}

Let us similarly define the complex electromagnetic vector 
\begin{equation}\label{eq:complemvec} 
\vec{\cal E} \equiv \vec{E}-i\,\vec{B}\,, 
\end{equation}
which in quantum mechanics represents photons with positive
helicity. Then the electromagnetic tensor
$F^{\,\alpha}_{\phantom{\alpha}\beta}$ of Eq.~(\ref{eq:fabmat}) can be
written as 
\begin{equation}\label{eq:compemtensor} 
\vec{F}\,=\,{\rm Re}\,(\,{\cal
  E}_k\,\vec{T}^{(k)} \,)\,. 
\end{equation}
Comparison with Eq.~(\ref{eq:etacompmat}) again brings out the
structural correspondence between the Mueller matrix and the
electromagnetic tensor, this time in a more concise and compact form. It also
shows how the electric and magnetic fields are inseparably linked, as
the real and imaginary parts of the same complex vector. 

It may be argued that the structural similarity between the Mueller
and electromagnetic formalisms is not unexpected, since the underlying
physics that governs the Mueller matrices is the electromagnetic
interactions between matter and radiation. The atomic transitions are
induced by the oscillating electromagnetic force of the ambient
radiation field when it interacts with the atomic electrons, and this
interaction is governed (in the classical description) by the force
law of Eq.~(\ref{eq:forcelaw}) with its electromagnetic tensor. This
is however not the whole story, since there are also profound
differences: As we will see in Sect.~\ref{sec:spin2} Stokes vectors and Mueller
matrices behave like spin-2 objects, while the electromagnetic tensor
is a spin-1 object. Another interesting aspect in the comparison
between Stokes vectors and the energy-momentum 4-vector is that
depolarization of Stokes vectors acts as if the corresponding 4-vector
has acquired ``mass'', as will be shown in Sect.~\ref{sec:depolmass}. Before we turn
to these topics we will in the next section show the correspondence
between the Mueller matrix and the Lorentz transformation matrix.

\section{Lorentz transformations and the Mueller absorption
  matrix}\label{sec:lorentz} 

Let $\vec{X}$ be the spacetime 4-vector: 
\begin{equation}\label{eq:xvect} 
\vec{X}\,\equiv\,\left(\matrix{\,\,ct\,\,\cr \,\,x\,\,\cr \,\,y\,\,\cr \,\,z\,\, }\right)\,\equiv\,\left(\matrix{\,\,x_0\,\,\cr \,\,x_1\,\,\cr \,\,x_2\,\,\cr \,\,x_3\,\, }\right)\,.
\end{equation}
 With the Lorentz transformation $\vec{\Lambda}$ we transfer to a new
system $\vec{X}^\prime$: 
\begin{equation}\label{eq:xprimex} 
\vec{X}^\prime =\vec{\Lambda}\,\vec{X}\,. 
\end{equation}
$\vec{\Lambda}$ represents rotations in Minkowski space, composed of
three spatial rotations $\phi_k$ and three boosts $\gamma_k$, which
may be regarded as imaginary rotations. Let us combine them into complex
rotation parameters $\alpha_k$ through 
\begin{equation}\label{eq:compalphagamphi} 
\alpha_k\,\equiv \,\gamma_k + i\,\phi_k\,. 
\end{equation}
Then the Lorentz transformation $\vec{\Lambda}$ can be written as 
\begin{equation}\label{eq:explorentz} 
\vec{\Lambda}\equiv e^{\vec{V}}\,, 
\end{equation}
where 
\begin{equation}\label{eq:vlormat} 
\vec{V}\,=\,{\rm Re}\,(\,\alpha_k\,\vec{T}^{(k)} \,)\,=\gamma_k\,
\vec{K}^{(k)}+\,\phi_k \,\vec{J}^{(k)}\,. 
\end{equation}
Explicitly, 
\begin{equation}\label{eq:lorentzmi}
\vec{V}\,= \left(\matrix{\,\,0&\phantom{-}\gamma_x&\phantom{-}\gamma_y&
\phantom{-}\gamma_z\,\,\cr \,\,\gamma_x&\phantom{-}0&
-\phi_z&\phantom{-}\phi_y\,\,\cr \,\,\gamma_y&\phantom{-}\phi_z&\phantom{-}0&-\phi_x\,\,\cr
\,\,\gamma_z&-\phi_y&\phantom{-}\phi_x& 
\phantom{-}0\,\, }\right)\,. 
\end{equation}
 Note that in quantum field theory a convention for the definition of
the $\vec{K}$ and $\vec{J}$ matrices that define the Lorentz algebra
is used, which differs by the
factor of the imaginary unit $i$ from the convention of Eq.~(\ref{eq:ikj}) used here, in
order to make $\vec{K}$ anti-hermitian and $\vec{J}$ hermitian
\citep{stenflo-zee2010}. 

Comparing with the Mueller matrix and the electromagnetic tensor we see the correspondence 
\begin{eqnarray}\label{eq:lorconnect} 
&\eta_k\,\,\longleftrightarrow \,\, \,\,\,\gamma_k\nonumber\,\,\,\longleftrightarrow \,\, \,\,E_k\\
&\,\,\rho_k\,\,\longleftrightarrow -\,  \phi_k\,\,\,\longleftrightarrow \,\, \,\,B_k\,.
\end{eqnarray}
The minus sign in front of $\phi_k$ is only due to the convention
adopted for defining the sense of rotations and is therefore
irrelevant for the following discussion of the physical meaning of the
structural correspondence. 

Relations (\ref{eq:lorconnect}) show how the Lorentz boosts
$\gamma_k$, which change the energy and momentum of the boosted object, relate to
both absorption $\eta_k$ and electric field $E_k$, while the spatial
rotations $\phi_k$, which do not affect the energy but change the
phase of the rotated object, relate to both dispersion or phase shift 
effects $\rho_k$ and to magnetic fields $B_k$.

\section{Analogy between depolarization and the emergence of
  mass}\label{sec:depolmass}
The structural correspondence between the $4\times 4$ Mueller
absorption matrix, the matrix representation of the covariant
electromagnetic tensor, and the Lorentz transformation matrix suggests
that there may be a deeper analogy or connection between the 4D Stokes vector
space and 4D spacetime. Let us therefore see what happens when we
introduce the Minkowski metric to the Stokes vector formalism. The
usual notation for the Minkowski metric is $\eta_{\mu\nu}$, but to
avoid confusion with the absorption matrix $\vec{\eta}$ that we have
been referring to in the present paper, we will use the notation
$\vec{g}$ or $g_{\mu\nu}$ that is generally reserved for a general
metric, but here we implicitly assume that we are only dealing with 
inertial frames, in which $g_{\mu\nu}=\eta_{\mu\nu}$. 

Assume that $\vec{I}_\nu =\vec{S}_\nu$ is the 4D Stokes vector, with
its transpose being $\vec{I}_\nu^T
=(I_\nu,\,Q_\nu,\,U_\nu,\,V_\nu)$. The scalar product in Minkowski
space is then 
\begin{equation}\label{eq:itetai}
\vec{I}_\nu^T
\vec{g}\,\vec{I}_\nu\,=\,I_\nu^2\,-\,(\,Q_\nu^2\,+\,U_\nu^2\,+\,V_\nu^2\,)\,,  
\end{equation}
which also represents the squared length of the Stokes vector in
Minkowski space. 

We know from polarization physics that the right-hand-side of
Eq.~(\ref{eq:itetai}) is always $\geq 0$, and equals zero only when the
light beam is 100\,\%\ (elliptically) polarized. Such fully polarized,
pure or coherent states are thus represented by null vectors, in
exactly the same way as the energy-momentum 4-vector $\vec{p}$ of
massless particles are also null vectors on the surface of null
cones. The energy-momentum vectors of massive particles live inside
the null cones. Similarly the Stokes vectors live inside and not on
the surface of null cones only if the light is not fully but partially
polarized. 

This comparison raises the question whether there is some deeper
connection between depolarization and the appearance of mass. In
polarization physics all individual (coherent) wave packages are
100\,\%\ polarized, and any coherent superposition of such wave
packages is also fully polarized. Partial polarization occurs
exclusively as a result of the {\it incoherent} superposition of
different, uncorrelated wave packages. In such cases it is customary to
represent the intensity $I_\nu$, which represents the energy or the
number of photons carried by the beam, as consisting of two parts, one
fraction $p_\nu$ that is fully polarized, and one fraction with
intensity $I_{\nu,\,u}$ that is unpolarized, with transposed Stokes
vector $I_{\nu,\,u}\,(1,\,0,\,0,\,0)$: 
\begin{equation}\label{eq:unpoldecomp}
I=p_\nu\,I \,+\,I_u \,,  
\end{equation}
where we have omitted index $\nu$ for simplicity except for $p_\nu$ (to
distinguish it from the momentum vector $p$ below). This fractional
polarization $p_\nu$ is  
\begin{equation}\label{eq:polfrac} 
p_\nu={I-I_u\over I}=\,{(Q^2 +U^2 +V^2)^{1/2}\over I}\,. 
\end{equation}

In comparison, in particle physics, the scalar product for the 4D
energy momentum vector $\vec{p}$ is 
\begin{equation}\label{eq:ptgp} 
\vec{p}^T\vec{g}\,\vec{p}\,=\,m^2\,c^2\,, 
\end{equation}
from which the well-known Dirac equation 
\begin{equation}\label{eq:ptetap} 
E^2=p^2\,c^2\,+\,m^2\,c^4 
\end{equation}
follows. While the emergence of the mass term corresponds to the
emergence of the unpolarized component $I_u$,
Eqs.~(\ref{eq:unpoldecomp}) and (\ref{eq:ptetap}) look different,
because the decomposition in Eq.~({\ref{eq:unpoldecomp}) has been done
  for the unsquared intensity $I$, while in Eq.~({\ref{eq:ptetap}) it
    is in terms of the squared components.  Since we have the freedom
    to choose different ways to mathematically decompose a quantity,
    this difference is not of particular physical significance. 

In current quantum field theories (QFT) the emergence of mass requires
the spontaneous breaking of the gauge symmetry, for which the Higgs
mechanism has been invented. It is postulated that all of space is
permeated by a ubiquitous Higgs field, which when interacting with the
field of a massless particle breaks the symmetry. When the particle
gets moved to the non-symmetric state it acquires mass. Because the phases
of the Higgs field and the field of the initially massless particle
are uncorrelated, the superposition of the fields is incoherent, which
may be seen as one reason for the breaking of the symmetry. 

In polarization physics the emergence of depolarization may also be
interpreted as a symmetry breaking, caused by the incoherent
superposition of different wave fields. Incoherence means that the
phases of the superposed fields are uncorrelated, which has the result that
the interference terms, all of which are needed to retain the symmetry,
vanish.

\section{Stokes vectors as spin-2 objects}\label{sec:spin2}
An object with spin $s$ varies with angle of rotation $\theta$ as
$s\,\theta$. For $s=\textstyle{1\over 2}$ one has to rotate $4\pi$
radians to return to the original state, for $s=2$ one only needs to
rotate $\pi$ radians,
and so on. Ordinary vectors, like the electric and magnetic fields
$\vec{E}$ and $\vec{B}$, rotate like spin-1 objects. It may
therefore come as a surprise that the Stokes vector rotates with twice
the angle, like a spin-2 object, in spite of the identical symmetry
properties of the Mueller matrix and the electromagnetic tensor. 

The resolution to this apparent paradox is found by distinguishing
between the kind of spaces in which the rotations are performed. In
the Minkowski-type space that is spanned by $I,Q,U,V$ as
coordinates, which is the Poincar\'e\ space in polarization physics
for a fixed and normalized intensity $I$, the transformation properties
are indeed those of a real vector, a spin-1 object. However, besides
Poincar\'e\ space the Stokes vector also lives in ordinary space, and
a rotation by $\theta$ of a vector in Poincar\'e\ space corresponds to
a rotation in ordinary space by $2\theta$. While being a spin-1 object
in Poincar\'e\ space, the same object becomes a spin-2 object in
ordinary space. 

The reason why it becomes a spin-2 object is that the Stokes vector
has substructure: it is formed from tensor products of Jones
vectors. Similarly Mueller matrices for coherent (100\,\%\ polarized)
wave packages are formed from tensor products of Jones matrices. While
the Jones vectors and matrices are spin-1 objects in ordinary space,
the bilinear products between them become spin-2 objects. 

The fundamental physics that governs the polarization physics does not
manifest itself  at the level of these spin-2 objects, because the basic
processes are the electromagnetic interactions between the radiation
field and the electrons (which may be bound in atoms), and these
interactions are described at the spin-1 level (since the
electromagnetic waves represent a spin-1 vector field). The Jones
matrices, or, in QM terminology, the Kramers-Heisenberg scattering
amplitudes, contain the fundamental physics. They are the basic
building blocks for the bilinear products, the spin-2 objects. 

This discussion points to the possibility that the physics of other
types of spin-2 objects, like the metric field in general relativity,
may be hidden, because the governing physics may take place within a spin-1
substructure level and would remain invisible if the spin-2 field
would be (incorrectly) 
perceived as fundamental, without substructure.

\section{Conclusions}\label{sec:conc}
Comparison between the Stokes formalism, the covariant formulation of
electromagnetism, and the Lorentz transformation shows that they all
share the same Lie algebra, namely the algebra of the Lorentz
group. This algebra is 6-dimensional (for instance in the case of
electromagnetism we have three electric field components + three magnetic
field components). While this is the algebra that is known to govern Lorentz
transformations and the related covariant formulation of
electromagnetism, it is not obvious why this group algebra should also apply
to the transformation of Stokes 4-vectors, which have been constructed
with the aim of being a powerful tool for the treatment of
partially polarized light. 

In spite of the common underlying group structure, there is also a
profound difference. While the electromagnetic field vectors and 
tensors are objects of a vector space with rotational properties of
spin-1 objects, Stokes vectors and Mueller matrices have the
rotational symmetries of spin-2 objects, because they are formed from
tensor products of spin-1 objects. This 
vector-field substructure contains the governing physics, where
everything is coherent, and where in the quantum description the probability
amplitudes or wave functions live and get linearly superposed to form mixed states with
certain phase relations. When we go to the spin-2 level by
forming bilinear products between the probability amplitudes, which
generates observable probabilities, or when we form bilinear products
between electric field vectors to generate 
quantities that represent energies or photon numbers, we get
statistical quantities (probabilities or energy packets) over which we
can form ensemble averages. If the phase relations of the mixed states in
the substructure are definite, we get interference effects and 100\,\%\
polarization for the ensemble averages, while if the phase relations
contain randomness (incoherent superposition) we get partial polarization. 

When comparing the Stokes 4-vector with the energy-momentum 4-vector
of a particle, 100\,\%\ elliptically polarized light corresponds to
massless particles, while the Stokes vector for partially polarized
light corresponds to the energy-momentum vector of a massive
particle. Depolarization thus has an effect as if the Stokes vector has
acquired ``mass'' by a symmetry breaking that is caused by the destruction of
coherences between mixed states.



\begin{thebibliography}{}
\expandafter\ifx\csname natexlab\endcsname\relax\def\natexlab#1{#1}\fi
\expandafter\ifx\csname url\endcsname\relax
  \def\url#1{\texttt{#1}}\fi
\expandafter\ifx\csname urlprefix\endcsname\relax\def\urlprefix{URL }\fi
\providecommand{\eprint}[2][]{\url{#2}}

\bibitem[{{Landi Degl'Innocenti} \& {Landi
  Degl'Innocenti}(1981)}]{stenflo-lanlan81}
{Landi Degl'Innocenti}, E., \& {Landi Degl'Innocenti}, M. 1981, Nuovo Cimento,
  62, B1

\bibitem[{{Landi Degl'Innocenti} \& {Landolfi}(2004)}]{stenflo-lanlan04}
{Landi Degl'Innocenti}, E., \& {Landolfi}, M. 2004, {Polarization in Spectral
  Lines}, vol. 307 of Astrophysics and Space Science Library (Kluwer)

\bibitem[{{Stenflo}(1994)}]{stenflo-book94}
{Stenflo}, J.~O. 1994, {Solar Magnetic Fields --- Polarized Radiation
  Diagnostics} (Kluwer)

\bibitem[{{Zee}(2010)}]{stenflo-zee2010}
{Zee}, A. 2010, {Quantum Field Theory in a Nutshell: Second Edition} (Princeton
  University Press)

\end{thebibliography}
\end{document}